\documentclass[fleqn,12pt]{wlscirep}
\usepackage[utf8]{inputenc}
\usepackage[T1]{fontenc}
\usepackage{multirow}

\title{CRAFTED - An exploratory database of simulated adsorption isotherms of metal-organic frameworks}

\author[1,2]{Felipe Lopes Oliveira}
\author[3]{Conor Cleeton}
\author[1,*]{Rodrigo Neumann Barros Ferreira}
\author[4]{Binquan Luan}
\author[3]{Amir H. Farmahini}
\author[3]{Lev Sarkisov}
\author[1]{Mathias Steiner}

\affil[1]{IBM Research, Av. República do Chile, 330, CEP 20031-170, Rio de Janeiro, RJ, Brazil.}
\affil[2]{Department of Organic Chemistry, Instituto de Química, Universidade Federal do Rio de Janeiro, Rio de Janeiro, RJ, Brazil.}
\affil[3]{Department of Chemical Engineering, Engineering A,  the University of Manchester, Manchester, M13 9PL, United Kingdom}
\affil[4]{IBM Research, 1101 Kitchawan Road, Yorktown Heights, 10519, NY, USA.}
\affil[*]{Corresponding author: rneumann@br.ibm.com.}

\begin{abstract}
    
    Grand Canonical Monte Carlo is an important method for performing molecular-level simulations and assisting the study and development of nanoporous materials for gas capture application. These simulations are based on the use of force fields and partial charges to model the interaction between the adsorbent molecules and the solid framework. The choice of the force field parameters and partial charges can significantly impact the results obtained, however, there are very few databases available to support a comprehensive impact evaluation. Here, we present a database of simulations of CO\textsubscript{2} and N\textsubscript{2} adsorption isotherms on 726 metal-organic frameworks taken from the CoRE MOF 2014 database. We performed simulations with two force fields (UFF and DREIDING), four partial charge schemes (no charge, Qeq, EQeq, and DDEC), and three temperatures (273, 298, 323 K). The resulting isotherms compose the \textbf{C}harge-dependent, \textbf{R}eproducible, \textbf{A}ccessible, \textbf{F}orcefield-dependent, and \textbf{T}emperature-dependent \textbf{E}xploratory \textbf{D}atabase (\textbf{CRAFTED}) of adsorption isotherms.

\end{abstract}
\begin{document}

\flushbottom
\maketitle

\thispagestyle{empty}


\section*{Background \& Summary}


    Carbon capture, storage, and utilization is considered as one of the key strategies required to reduce anthropogenic carbon dioxide emissions and their impacts on climate change~\cite{mac2017role}. The most viable option for this approach is to focus on CO\textsubscript{2} capturing from the point sources, such as fossil fuel power plants, fuel processing plants and other industrial plants where carbon capture technology can be applied to streams with the industrial scale flow rates\cite{metz2005ipcc}. However, despite several decades of intensive research, carbon capture in an economically viable way remains an enormous challenge~\cite{sholl2016seven} . 
    
    Adsorption processes are considered to be a promising alternative to the conventional absorption processes due the their low regeneration energy, high selectivity, and high capture capacity~\cite{samanta2012post}. Combined, these characteristics may lead to energy-efficient processes for industrial scale  capture and utilization of greenhouse gases (GHG). At the heart of a typical adsorption process for gas separation, such as the Pressure Swing Adsorption process, is the active adsorbent material; and the efficiency of the process crucially depends on the properties of this material.
    Within the different adsorbent materials that are potentially available for this process, crystalline nanoporous materials such as metal-organic frameworks (MOF)~\cite{yaghi2003reticular, furukawa2013chemistry, chen2022reticular, maia2021co}, covalent organic frameworks (COF)~\cite{cote2005porous, diercks2017atom, maia2021co2}, zeolitic imidazolate frameworks (ZIF)~\cite{banerjee2008high, yang2017principles, wang2022large}, and zeolites~\cite{lin2012silico} feature many of the necessary characteristics for a solid sorbent for efficient gas separation under the conditions of interest.
    
    These families of materials contain hundreds of thousands synthesized structures and countless more hypothetical ones, featuring pores of different size, shape, and chemical characteristics. This creates large exploration space for studies that seek to identify the best candidates for a given gas capture application. This endeavour, however, is not possible via a brute-force experimental campaign.
    The number of large databases built upon experimental~\cite{chung2014computation, tong2017exploring, moghadam2017development, chung2019advances, ongari2019building} and hypothetical~\cite{colon2017topologically, rosen2021machine} structures, combined with the continuous growth of diversity and scope of new materials due to the advancements in digital reticular chemistry,~\cite{lyu2020digital, ji2021molecules} make high-throughput computational screening (HTCS) approaches an an imperative strategy for efficient exploration of the vast chemical landscape of crystalline nanoporous adsorbents~\cite{colon2014high, majumdar2021diversifying}. 
    
    Most of the HTCS studies for carbon capture and related problems are based on using Grand Canonical Monte Carlo (GCMC) simulations to generate adsorption data. This data is then used to form some simple material performance metrics or is passed on to the process level modelling to explore performance of candidate materials under the realistic process conditions. 
    
    To perform molecular simulations, such as GCMC, one needs a set of parameters that describe the interactions among the adsorbate molecules, and between the adsorbate molecules and the atoms of the adsorbent material; this set of parameters is called a force field. 
    
    Over the years of the development, many force fields have been developed for various purposes and several options are available to describe adsorption of gases such as carbon dioxide in materials such as MOFs. Invariably, the predicted equilibrium adsorption data and, consequently ranking of the materials and the recommendations of the screening study will depend on the choice of the force field. 
    
    This poses several fundamental questions. How sensitive is the adsorption data to the choice of the force field parameters? How does this sensitivity vary across different categories or classes of materials? And ultimately, is a ranking of porous materials for a particular application a robust result or it is contingent on using a particular force field? 
    
    To start to explore these questions one needs a representative mass of adsorption data covering typical choices of the force field parameters, materials, gases and conditions. This defines the remit of the current article where we tasked ourselves with building such a database (or at least, the first block of conditions). 
    
    To explain the contents of the database and our approach, let us delve first into components of the classical force fields and the typical options available for the studies of adsorption of gases in MOFs and related materials. In the classical force fields, the non-bonded interactions are modeled as a sum of van der Waals and Coulomb potentials~\cite{dubbeldam2019design}. The van der Waals interactions between the adsorbed molecules and the framework are usually modeled by the Lennard-Jones (LJ) potential, which is an effective potential with two fitted parameters that can capture most of the intermolecular effects relevant to physisorption. The parameters for the atoms can be taken from the generic force fields such as the Universal Force Field (UFF)~\cite{rappe1992uff}, DREIDING~\cite{mayo1990dreiding} and TraPPE~\cite{potoff2001vapor}, with the interactions between different atom species computed using mixing rules such as Lorentz-Berthelot~\cite{allen2017computer} or Jorgensen~\cite{jorgensen1986optimized}.

    The Coulombic interactions are modeled by partial atomic charges assigned to the atoms which need to be calculated for each material. There are several charge assignment methods available, and they can be divided into two main groups: \textit{i.} methods derived from quantum chemistry calculations (e.g. RESP~\cite{bayly1993well}, CHELP~\cite{breneman1990determining}, REPEAT~\cite{campana2009electrostatic}, and DDEC~\cite{manz2016introducing, limas2016introducing, manz2017introducing}) and \textit{ii.} methods based on charge equilibration (e.g. Qeq~\cite{rappe1991charge}, PQeq~\cite{ramachandran1996toward}, EQeq~\cite{wilmer2012extended}, and FC-Qeq~\cite{wells2015charge}). Although there is some consensus that the approaches such as DDEC (based on electronic structure calculations) are more accurate, methods such as EQeq can present sufficiently good results that, combined with their low computational cost, makes them attractive choices for HTCS studies~\cite{altintas2020role, liu2022benchmarking}.
    
    Lately, there have been several studies evaluating the accuracy of different methods for calculating partial atomic charges~\cite{zheng2009computational, hamad2015atomic, ongari2018evaluating, sladekova2020effect, liu2022benchmarking}, however, little is known about the combined impact of force field and partial charge selection on material-level analysis and its implication on process-level performance metrics. Furthermore, the parameters of force fields such as UFF and DREIDING were fitted employing specific partial charge schemes (Gasteiger~\cite{gasteiger1980iterative} for DREIDING and Qeq~\cite{rappe1992uff} for UFF), thus the combination of these parameters with different charge assignment methodologies, even if more accurate, may not necessarily generate better results.
    
    These considerations guide us on the choices of the parameters of the force fields to consider in the database. 
    
    The database contains simulated adsorption isotherms for 726 MOFs selected from the CoRE MOF 2014~\cite{chung2014computation} database. The simulations were performed for the adsorption of CO\textsubscript{2} and N\textsubscript{2} with two force fields (UFF and DREIDING), four partial charge schemes (no charge, Qeq, EQeq, and DDEC), at three temperatures (273, 298, 323 K). The resulting isotherms compose the \textbf{C}harge-dependent, \textbf{R}eproducible, \textbf{A}ccessible, \textbf{F}orcefield-dependent, and \textbf{T}emperature-dependent \textbf{E}xploratory \textbf{D}atabase (\textbf{CRAFTED}) of adsorption isotherms. CRAFTED provides a convenient platform to explore the sensitivity of simulation outcomes to molecular modeling choices at the material (structure-property relationship) and process levels (structure-property-performance relationship).

\section*{Methods}

    

\subsection*{Structure selection} \label{sec:selection}

    The 2932 structures present in the CoRE MOF 2014~\cite{chung2014computation} database were analyzed and 726 structures were retained in our analysis. This subset of 726 comprises all materials from the CoRE MOF 2014 database for which all atom types are present in both DREIDING and UFF force fields. Throughout this work, this subset of structures will be referred to as ``CRAFTED structures''.
    
\subsection*{Partial charges calculation}

    The DDEC partial charges~\cite{manz2016introducing, limas2016introducing, manz2017introducing} were taken without modification from the CoRE MOF 2014~\cite{chung2014computation} database. The EQeq partial charges~\cite{wilmer2012extended} were calculated using the the extended charge equilibration method as implemented in the EQeq software~\cite{eqeq} v1.1.0. The Qeq partial charges~\cite{rappe1991charge} were calculated using the default implementation available in RASPA~\cite{dubbeldam2016raspa}.

\subsection*{Grand Canonical Monte Carlo simulations}

    Atomistic Grand Canonical Monte Carlo (GCMC) simulations were performed using a force field-based algorithm as implemented in RASPA~\cite{dubbeldam2013inner, dubbeldam2016raspa} v2.0.45. Interaction energies between non-bonded atoms were computed through a combination of Lennard-Jones (LJ) and Coulomb potentials
    
    \begin{equation}
        U_{ij}(r_{ij}) = 4\varepsilon_{ij} \left [  \left( \frac{\sigma_{ij}}{r_{ij}} \right)^{12}  - \left( \frac{\sigma_{ij}}{r_{ij}} \right)^6\right] + \frac{1}{4\pi \epsilon} \frac{q_iq_j}{r_{ij}}
    \end{equation}
    
    where $i$ and $j$ are interacting atom indexes and $r_{ij}$ is their interatomic distance. $\varepsilon_{ij}$ and $\sigma_{ij}$ are the well depth and diameter, respectively. The LJ parameters between atoms of different types were calculated using the Lorentz-Berthelot mixing rules
    
    \begin{equation}
        \varepsilon_{ij} = \sqrt{\varepsilon_{ii}\varepsilon_{jj}}, \qquad \sigma_{ij} = \frac{\sigma_{ii} + \sigma_{jj}}{2} 
    \end{equation}
    
    LJ parameters for framework atoms were taken from Universal Force Field (UFF)~\cite{rappe1992uff} or DREIDING~\cite{mayo1990dreiding} (see \autoref{tab:forcefields}). The parameters for the adsorbed molecules were taken from the TraPPE~\cite{potoff2001vapor} force field (see \autoref{tab:trappe}). All simulations were performed with 10,000 Monte Carlo cycles. Swap (insertion or deletion with with a probability of 50\% for each), translations, rotations, and re-insertions moves were tried with probabilities 0.5, 0.3, 0.1, and 0.1, respectively. To avoid the use of long initialization cycles, each isotherm was calculated in a single simulation, with each pressure point of the simulation starting from the result of the previous one. The uptake values for each pressure were obtained by averaging over the GCMC equilibrium phase, determined using the Marginal Standard Error Rule. For more information, please refer to section \textit{Automatic transient regime detection and truncation}.

    All atoms in the MOF were held fixed at their crystallographic positions. The number of unit cells used was different for each MOF to ensure that the perpendicular lengths of the supercell were greater than twice the cutoff used. The cutoff for Lennard-Jones and charge-charge short-range interactions was 12.8 \AA{} and the Ewald sum technique was applied to compute the long-range electrostatic interactions with a relative precision of 10\textsuperscript{–6}. The Lennard-Jones potential was shifted to zero at the cutoff. Fugacities needed to impose equilibrium between the system and the external ideal gas reservoir at each pressure were calculated using the Peng-Robinson equation of state~\cite{peng1976industrial} with the critical parameters for each gas taken from \autoref{tab:critical}. All GCMC uptake data report the absolute adsorption value in mol/kg units.
    
    The enthalpy of adsorption was computed as
    
    \begin{equation}
        \Delta H = \frac{\langle U \cdot N\rangle - \langle U \rangle \langle N\rangle}{\langle N^2\rangle - \langle N\rangle^2 } - RT
    \end{equation}
    
    where $N$ is the number of adsorbates on the simulation box and $U$ is the potential energy\cite{vlugt2008computing}. All the adsorption enthalpy values are reported in kJ/mol and are the values as calculated by RASPA without further modification. 

\subsection*{Lennard-Jones parameters}

    The Lennard-Jones parameters for DREIDING and UFF force fields used in the calculations for the framework atoms are shown in \autoref{tab:forcefields}. For simplicity, only the atoms that are present in both UFF and DREIDING are shown. The TraPPE parameters used for the gas molecules are present in \autoref{tab:trappe}. The critical parameters used in the Peng-Robinson equation to calculate the fugacity are present in \autoref{tab:critical}.

\subsection*{Automatic equilibration detection and truncation}

    To eliminate the use of long initialization cycles, the Marginal Standard Error Rule (MSER)~\cite{white1997effective} was applied to automatically detect the ideal truncation point using the pyMSER package v1.0.12~\cite{pymser}, so that the averages were taken only over the equilibrated phase of the simulation. The output of this method is the equilibrated average of the observable, alongside an uncertainty metric. Here we used the uncorrelated standard deviation, as explained in the next sub-section.
    
    The MSER defines the start of the equilibrated region $\hat{d}(n)$ by solving the minimization problem:

    \begin{equation}
        \hat{d}(n) =  \underset{0 \leq k \leq n-2}{\mathrm{arg\,min}} \, g_n(k) \qquad \text{where} \qquad g_n(k) = \frac{1}{(n - k)^2} \sum_{j=k}^{n-1}(Y_j - \bar{Y}_{n,k})^2 = \frac{S^2_{n,k}}{n-k}
    \end{equation}
    
    The Left-most Local Mininum (LLM) version of MSER was used in a batched data with batch size of 5.

\subsection*{Uncorrelated Standard Deviation (uSD)}

    To use an uncertainty metric that reflects, at the same time, the real dispersion of the simulated values and the number of cycles used for this simulation, the \textit{uncorrelated standard deviation} (uSD) was used as an uncertainty metric. To calculate this quantity, first the number of uncorrelated states in the simulation is estimated by calculating the \textit{autocorrelation time}. The equilibrated data is divided into chunks so that each chunk has a number of data points equivalent to the \textit{autocorrelation time}. Then, the average value of each chunk is calculated and the standard deviation over this list of uncorrelated average values is calculated as the uSD.
    
    The autocorrelation time is estimated by calculating the autocorrelation function of the equilibrated data using the method implemented in the \texttt{numpy} library.~\cite{harris2020array} An exponential decay function is fitted over the values of the autocorrelation function and the autocorrelation time is calculated as the half-life of this exponential decay.

\subsection*{Automatic simulations workflow}

\begin{figure}[ht]
        \centering
        \includegraphics[width=.8\textwidth]{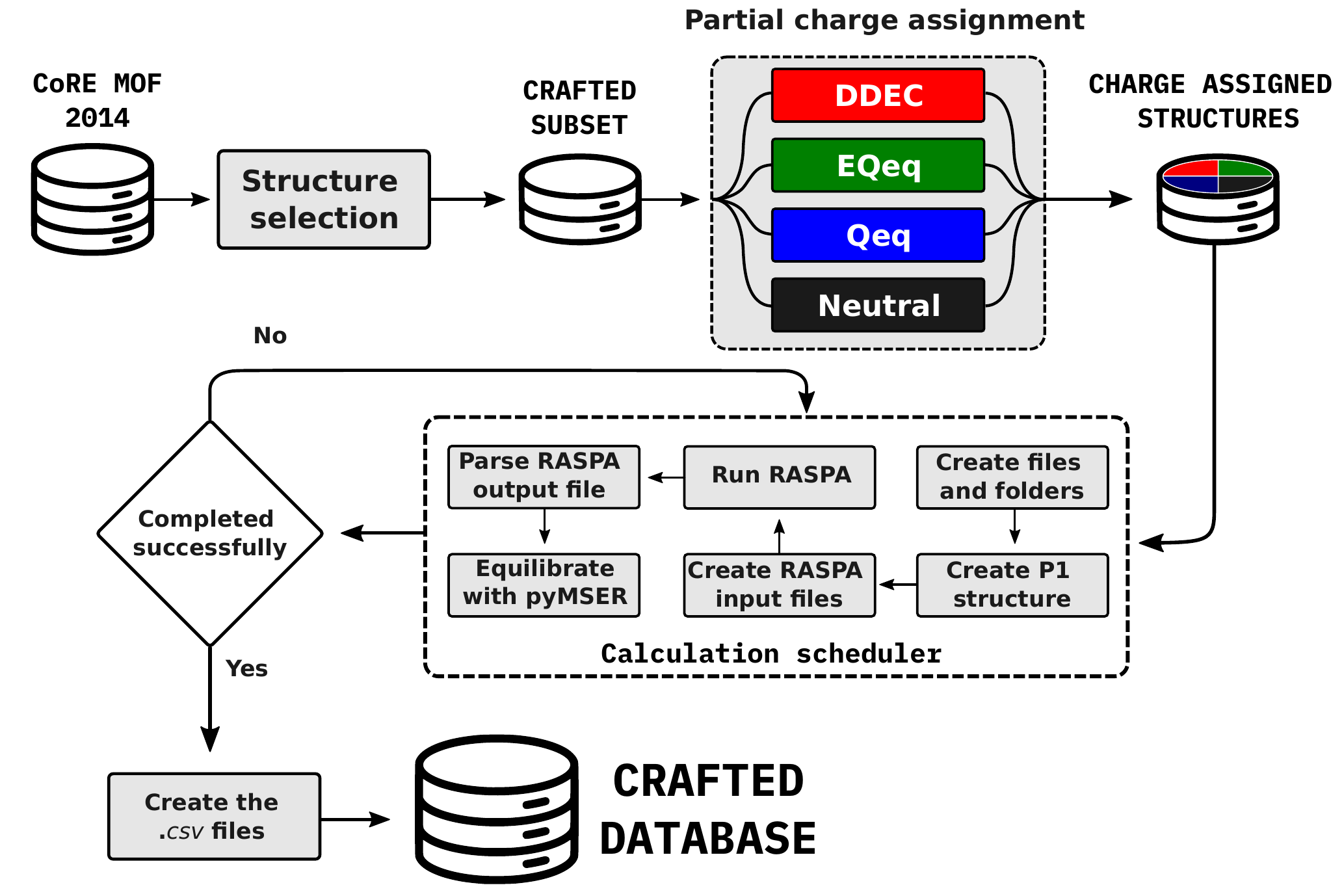}
        \caption{Schematic representation of the main steps in the automated simulation workflow.}
        \label{workflow}
    \end{figure}

    A set of scripts composed of three stages were created to automate the isotherm generation process. First, a pre-processing step is performed where partial charges are calculated for all structures. Next, a set of calculation scheduler scripts are executed, where steps such as copying the force field and CIF files, creating a supercell with P1 symmetry, writing the RASPA input file, running the RASPA simulation, parsing the RASPA output, and performing the MSER analysis of the results for averaging over the equilibrated phase of the simulation, are run in sequence. Finally, a post-processing script is executed to analyze the results, resubmit incomplete calculations and create the isotherm database. A simplified scheme containing the main steps of this workflow is present on \autoref{workflow}.

\subsection*{Revised Autocorrelations (RACs) descriptors}

    To understand the diversity of our subset of CRAFTED MOFs, and understand how representative they are with respect to the MOF material class, revised autocorrelations (RACs) descriptors were calculated using the molSimplify software v1.7.1~\cite{ioannidis2016molsimplify}. RACs are built by generating a crystal graph derived from the adjacency matrix computed for the primitive cell of the crystal structure and calculating the discrete correlations between atomic properties (Pauling electronegativity, nuclear charge, etc.) over the atoms. The correlations are composed of the products (\autoref{prod}) and the differences (\autoref{diff}) of an atomic property $P$ for atom $i$, which is selected from the $start$ atom list and is correlated to atom $j$ selected from the $scope$ atom list when they are separated by $d$ number of bonds.
    
    \begin{equation}
        ^{start}_{scope}P^{prod}_d = \sum^{start}_i \sum^{scope}_j P_i P_j\delta(d_{i,j}, d)
        \label{prod}
    \end{equation}

    \begin{equation}
        ^{start}_{scope}P^{diff}_d = \sum^{start}_i \sum^{scope}_j (P_i - P_j)\delta(d_{i,j}, d)
        \label{diff}
    \end{equation}

    Six atomic properties were used to compute RACs: atom identity (I), connectivity (T), Pauling electronegativity ($\chi$), covalent radii (S), nuclear charge (Z) and polarisability ($\alpha$). These properties are used to generate metal-centred, linker and functional-group descriptors. To generate a fixed length descriptor, the averages of these descriptors were used, thus generating 156 features (40 for metal chemistry, 68 for linker chemistry and 48 for functional group chemistry) for each MOF structure.
    
\subsection*{Dimensionality reduction and cluster analysis}
     
    To reduce the dimensionality of the feature vectors that describe CRAFTED structures, the t-Stochastic Neighbor Embedding (t-SNE)~\cite{van2008visualizing} method was employed. Different fitting parameters where used for each set of descriptors for metal chemistry ($perplexity$ = 45, $early\_exaggeration$ = 1, $learning rate$ = 50), linker chemistry ($perplexity$ = 100, $early\_exaggeration$ = 1, $learning rate$ = 50), functional groups chemistry ($perplexity$ = 50, $early\_exaggeration$ = 1, $learning rate$ = 200) and geometric properties ($perplexity$ = 50, $early\_exaggeration$ = 1, $learning rate$ = 200). 
     
    For the t-SNE projections of the geometric properties, 14 features were used: largest included sphere (D\textsubscript{is}), the largest free sphere (D\textsubscript{fs}), largest included sphere along a free path (D\textsubscript{isfs}), volumetric accessible area (ASA\textsubscript{m2/cm3}), gravimetric accessible area (ASA\textsubscript{m2/g}), volumetric non-accessible area (NASA\textsubscript{m2/cm3}), gravimetric non-accessible area (NASA\textsubscript{m2/g}), unit cell volume, crystal density, accessible volume fraction (AVF), non-accessible volume fraction (NAVF), accessible volume (AV\textsubscript{cm3/g}), non-accessible volume (NAV\textsubscript{cm3/g}), and the number of pockets (n\textsubscript{pockets}). All these properties were calculated using Zeo++ v0.3.~\cite{willems2012algorithms} The chemical descriptors used for the t-SNE projections of the MOF structures were described in the previous section. All structures that could not have their descriptors calculated were removed from the list.
    
    For the unsupervised cluster analysis, the Density-Based Spatial Clustering of Applications with Noise (DBSCAN)~\cite{bi2012dbscan} method was employed. The DBSCAN analysis was performed using the standardized and scaled set of descriptors with $eps$ = 0.09 and minimum number of samples per cluster of 25.

\section*{Data Records}

    
    
    CRAFTED provides 34,848 isotherm files and 34,848 adsorption enthalpy files resulting from the GCMC simulation of two gases (CO\textsubscript{2} and N\textsubscript{2}) on 726 MOF structures at three temperatures (273, 298,  and 323 K) using two force fields (UFF and DREIDING) and 4 partial charge methods (no charges, Qeq, EQeq, and DDEC). Alongside the isotherm data, the charge-assigned CIF files, force field and molecule definition files are provided, to ensure reproducibility and facilitate a future database expansion. 
    
    Each isotherm file corresponds to a comma-separated value (CSV) file containing three labeled columns corresponding to pressure (in Pa), uptake volume and its uncertainty (in mol/kg).  The file names follow the pattern \texttt{Q\_MOF\_FF\_GAS\_T.csv}, therefore the isotherm file named \texttt{DDEC\_ABUWOJ\_UFF\_CO2\_273.csv} contains the data corresponding to the CO\textsubscript{2} adsorption isotherm at 273K on the ABUWOJ MOF with DDEC partial charges using the UFF forcefield. The adsorption enthalpy file names follows the same pattern.
    
    The adsorption enthalpy files correspond to a CSV file ontaining three labeled columns corresponding to pressure (in Pa), adsorption enthalpy and its uncertainty (in kJ/mol), following the same naming pattern as the isotherm files. 

    The RASPA input file names follows the same pattern as the isotherm files (\texttt{Q\_MOF\_FF\_GAS\_T.input}), the cif files are separated into folders according to their partial charge (Qeq, EQeq, DDEC, and NEUTRAL) type and the force field files are separated into folder according to their type (UFF and DREIDING). 
    
    All data are available in a dedicated Zenodo repository at \url{https://doi.org/10.5281/zenodo.7106174}.
    
\section*{Technical Validation}
    
    \subsubsection*{Chemical and geometrical diversity of CRAFTED subset of structures}
    
    The selection of a subset of structures that can be modeled simultaneously by both UFF and DREIDING force fields may impose a limitation on the structural and chemical representativeness of the CRAFTED database and, consequently, on the results obtained with this data. To ensure that CRAFTED MOFs form a group that represents the great diversity of experimentally realised MOFs, both the geometrical and chemical diversities must be represented.

    To evaluate the geometrical diversity, the pore size (such as the largest included sphere, largest free sphere and largest included sphere along a free path), void fraction, density, unit cell volume, specific area (both gravimetric and volumetric), pore volume (both gravimetric and volumetric) and the number of pockets (non accessible pores) were used as the descriptors for t-SNE projection. Both the accessible and non-accessible specific area and pore volume was used. 

    For the chemical diversity, the revised autocorrelations (RACs) descriptors~\cite{janet2017resolving} were used. This approach has been successfully applied to study transition metal chemistry~\cite{nandy2018strategies} and the chemical diversity of MOF datasets~\cite{moosavi2020understanding}. The chemical characteristics of the MOFs were divided into three categories: metal node chemistry, organic linker chemistry and functional groups chemistry.  

\begin{figure}[ht]
        \centering
        \includegraphics[width=\linewidth]{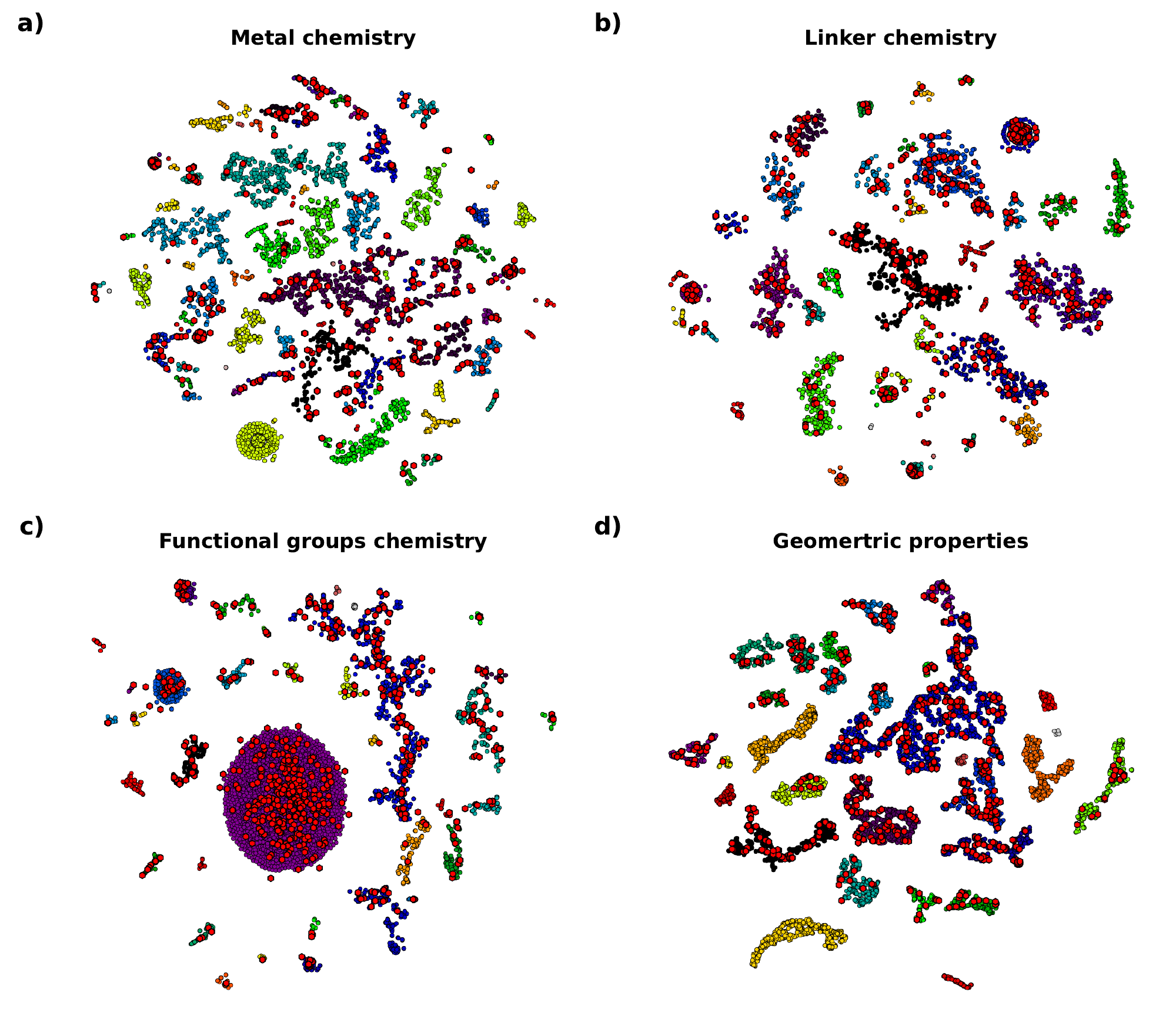}
        \caption{t-SNE representation of the CRAFTED structures (red points) and the CoRE MOF 2019 (coloured points) database on different domains of MOF chemistry based on RACs descriptors and geometric properties. (a) metal chemistry, (b) linker chemistry, (c) functional groups chemistry, (d) geometric properties. The color scheme correspond to the cluster assigned by DBSCAN}
        \label{fig:tsne}
    \end{figure}

    \autoref{fig:tsne} shows the t-SNE projection onto 2D maps of the four selected groups of descriptors for the structures in the CoRE MOF 2019 database (colored circles) and the selected structures for CRAFTED database (red hexagons). Although CRAFTED structures were taken from the first version of CoRE MOF from 2014, here the comparison is made with the second version of this database, from 2019, as it has a greater number and diversity of structures.

    To numerically evaluate the overlap of the databases in t-SNE projected space, the DBSCAN method was used. This method is an unsupervised machine learning technique used to identify clusters of varying shape and size, grouping points that are close to each other. Since the t-SNE method reduces the feature space by modelling structures with similar features as nearby points and distinct features as distant points, the groups found by the DBSCAN method will share similarities within the original feature space. 
    
    The limitation imposed by the DREIDING force field reduces the diversity of metal chemistry observed in the CRAFTED database compared to the structures presented in CoRE MOF 2019, as shown in \autoref{fig:tsne}(a). However, 59 of the 95 clusters found (~62\%) have some structure present in CRAFTED, indicating that even with a limited metal cluster composition, the CRAFTED structures show a good representation of the chemical diversity present in CoRE MOF 2019.
    
    The chemical diversity of both organic linker and functional groups is much better represented within the CRAFTED structures, as can be seen in \autoref{fig:tsne}(b) and \autoref{fig:tsne}(c). In both cases, one can see that the points from both CRAFTED and CoRE MOF 2019 structures are equally dispersed across 2D space. Additionally, 36 of the 40 clusters identified for linker chemistry (90\%) and 31 of the 35 clusters identified for functional group chemistry (~89\%) contain structures present in CRAFTED.

\begin{figure}[ht]
        \centering
        \includegraphics[width=\linewidth]{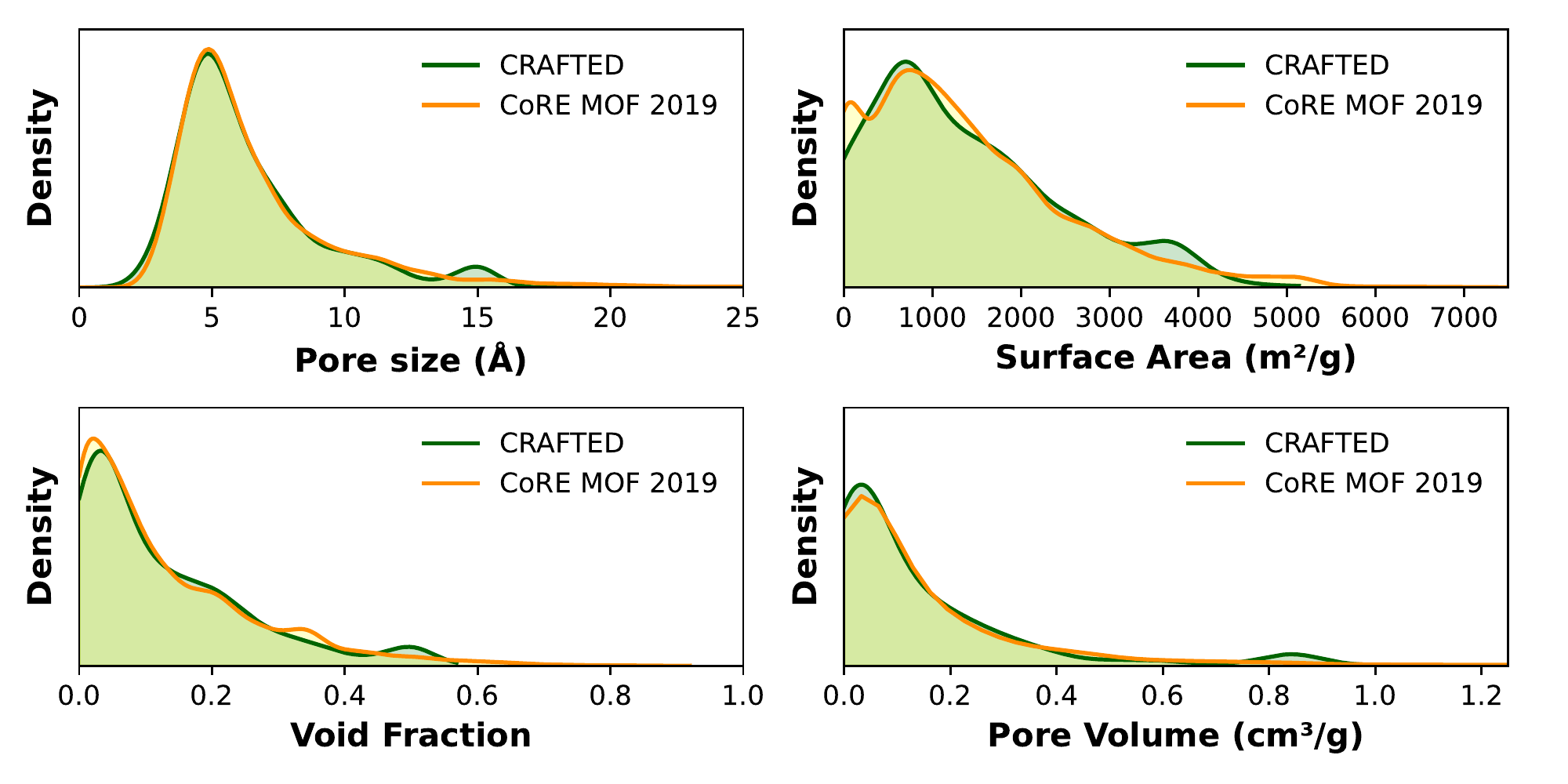}
        \caption{Comparison between the main geometric properties of the structures present in CRAFTED and CoRE MOF 2019.}
        \label{fig:geometric}
    \end{figure}

    The geometric properties are also well represented by the CRAFTED database, as shown in \autoref{fig:tsne}(d). From the 28 clusters identified, 21 (75\%) present structures from CRAFTED. Additionally, \autoref{fig:geometric} shows the distribution density of the values for the main geometric properties presented by the structures on CRAFTED and CoRE MOF 2019. One sees that both databases contain similar distributions, showing that even with a smaller number of structures, the CRAFTED database is exemplary of synthesized MOF structural properties.
    
    \subsubsection*{General impact of force field and partial charge selection}
    
\begin{figure}[ht]
        \centering
        \includegraphics[width=1\textwidth]{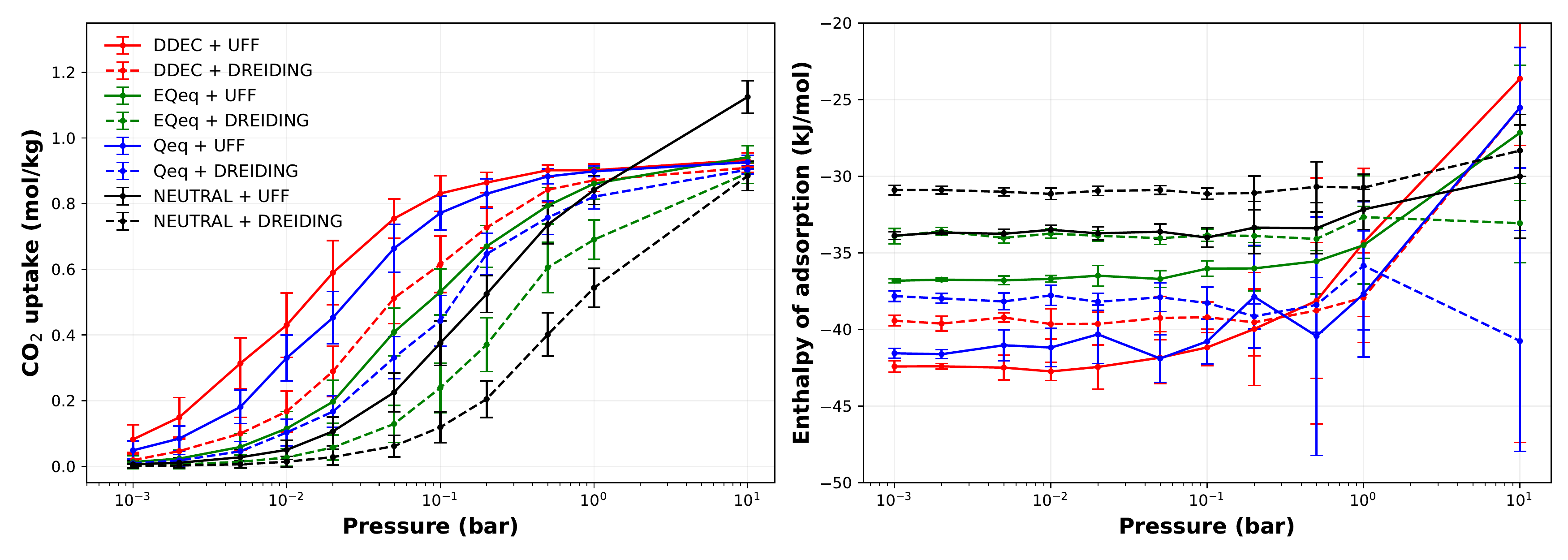}
        \caption{Example of the impact of force field and partial charge choice on the simulated adsorption isotherms of CO\textsubscript{2} on BONWIL MOF at 273K.}
        \label{fig:impact}
    \end{figure}

     To illustrate the impact of molecular-level simulation parameters (force field and partial charge) on the outcome of the GCMC simulations, we show in \autoref{fig:impact} the absolute uptake and enthalpy of adsorption of CO\textsubscript{2} on the BONWID MOF at 237K. At 0.1 bar, a typical pressure used in adsorption-based pressure swing adsorption (PSA) processes for CO\textsubscript{2} capture\cite{park2019well}, the uptake values range from 0.12 to 0.83 mol/kg and every combination of partial charge and force field yields a different value, while at 10 bar almost all conditions resulted in similar uptake. This dispersion of results is also reflected on the enthalpy of adsorption that ranges from -31 to -43 kJ/mol and are fairly different for every combination of parameters.
     
\begin{figure}[ht]
        \centering
        \includegraphics[width=1\textwidth]{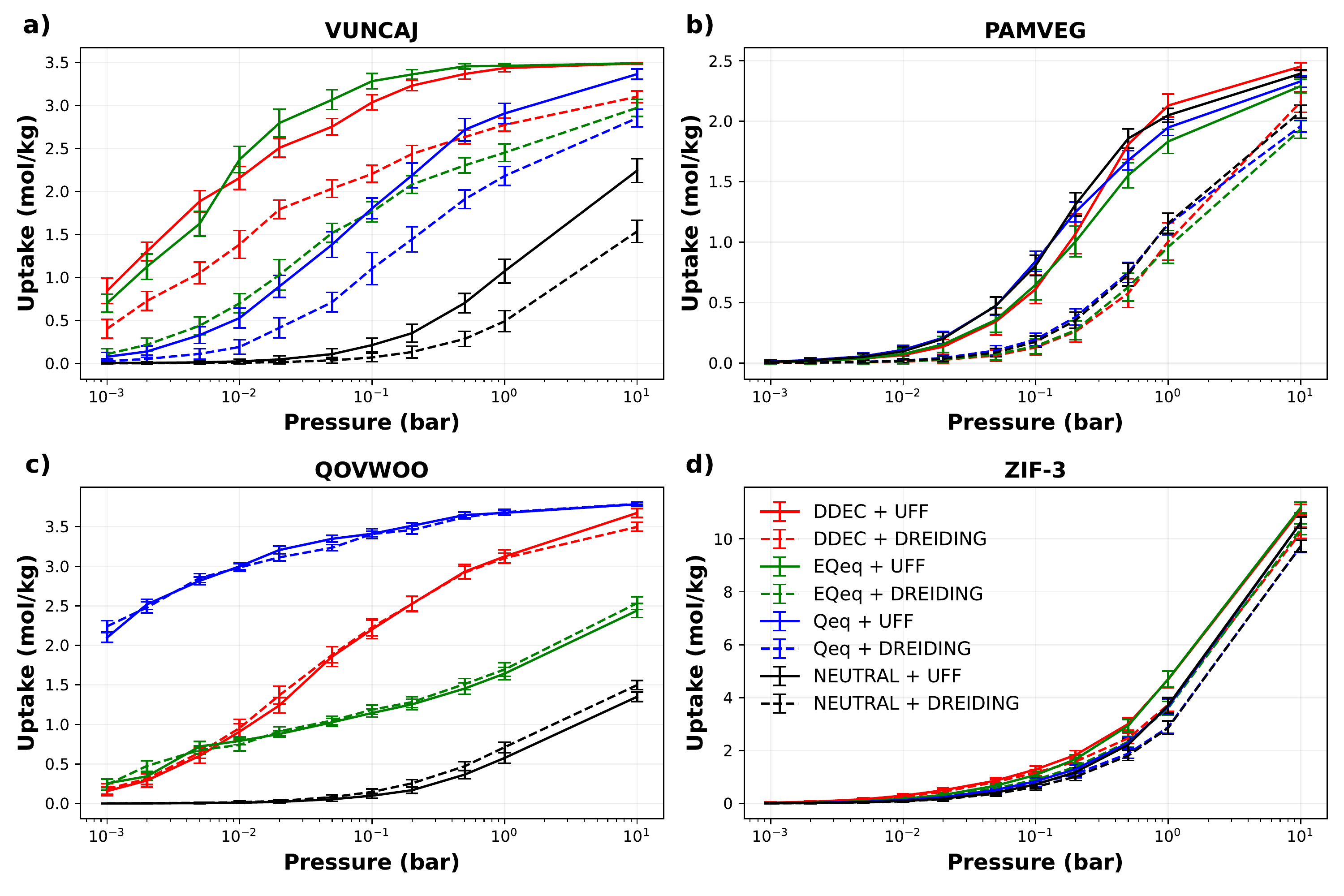}
        \caption{Examples of four representative behaviours found in the dataset: (a) high sensitivity to force field and partial charge, (b) high sensitivity to force field and low sensitivity to partial charge, (c) low sensitivity to force field and high sensitivity to partial charge, and (d) low sensitivity to force field and partial charge.}
        \label{fig:fourcases}
    \end{figure}

     Among the CRAFTED materials, one finds a diversity of responses to the choices of force field and partial charge method. Four representative cases are highlighted in \autoref{fig:fourcases}. For some materials, the uptake is highly dependent on both the force field and partial charge, others are only sensitive to one of these parameters, and some are not sensitive to any. Therefore it is possible to anticipate that most studies that depend directly on the results of these molecular simulations, such as high-throughput computational screening or multiscale processes modelling, may also present different degrees of dependence on the choice of parameters.

\section*{Usage Notes}

    
    The CRAFTED database can be found on Zenodo \cite{lopes_oliveira_felipe_2022_7106174} repository. The database contains 34,848 CSV files with the isotherm adsorption data (pressure, uptake, and uncertainty), definition files for UFF and DREIDING force fields, 2,904 CIF files for all 726 structures with all 4 partial charges considered (NEUTRAL, DDEC, EQeq, and Qeq). The database also contains 34,848 files containing the adsorption enthalpy. All 34,848 RASPA input files to facilitate the reproduction of the isotherm simulations and 2 CSV files containing the set of RAC and geometrical descriptors calculated with molSimplify and Zeo++, respectively.
    
\begin{figure}[ht]
        \centering
        \includegraphics[width=\linewidth]{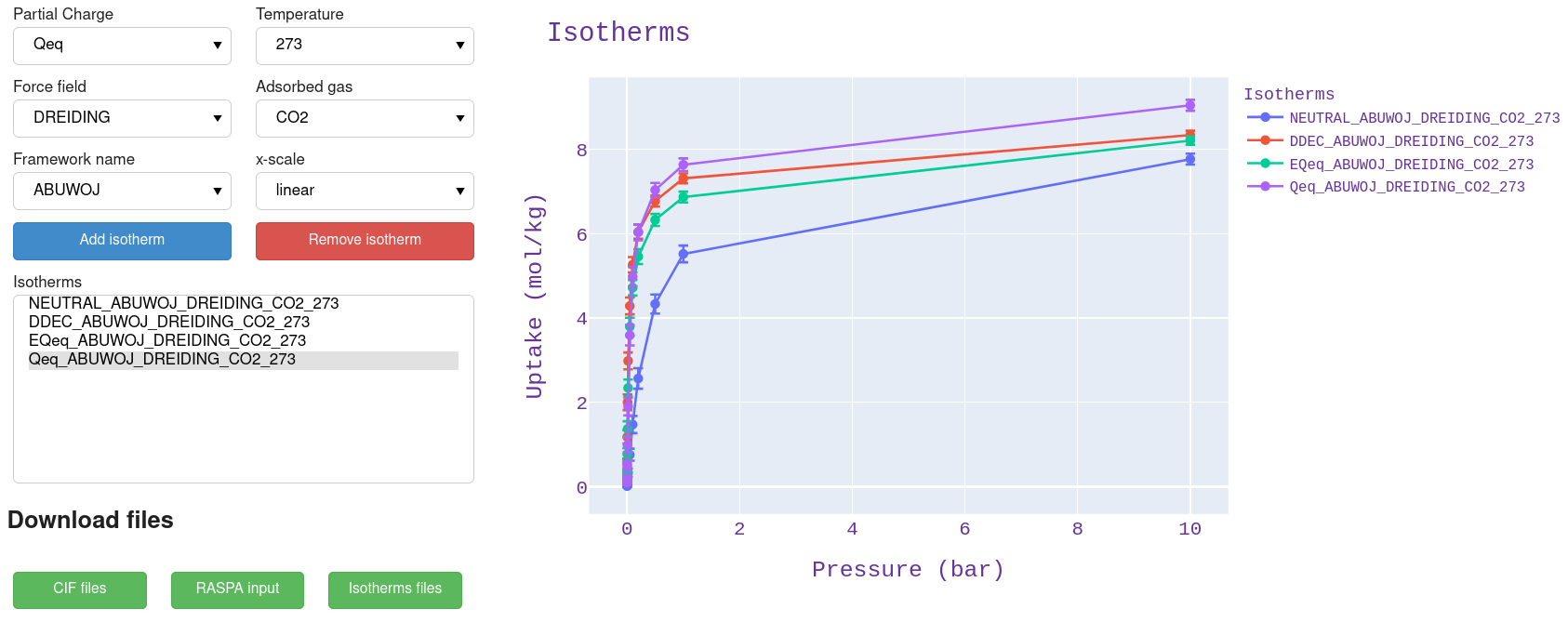}
        \caption{Screenshot of the interactive interface for isotherm visualization.}
        \label{fig:panel}
    \end{figure}

    To facilitate the exploration and visualization of the isotherms present on CRAFTED, we also developed an interactive visualization interface based on a Jupyter notebook and \texttt{panel}. This interface allows the user to select a set of specific conditions --- e.g. partial charge, temperature, force field, adsorbed gas, and material name --- for each isotherm, thus facilitating a quick and easy visual comparison of the data. In addition, it is possible to download the CIF files, the inputs for the GCMC simulation with RASPA, and the data of the selected isotherms, thus facilitating the reproduction of the data present in CRAFTED. An example of the interface is shown in \autoref{fig:panel}. We recommend the user to set up a Python environment using the \texttt{environment.yml} or \texttt{requirements.txt} provided therein.
    
    A Jupyter notebook with the code to perform the t-NSE dimensionality reduction and the DBSCAN unsupervised clusterization analysis with the necessary files containing the RAC and geometric descriptors for the CoRE MOF 2019 is also provided alongside the CRAFTED data, providing an easy reproduction of the results presented in \autoref{fig:tsne}.
    
    Finally, we would like to highlight some points to show why this database benefits the scientific community. Machine learning (ML) and data-driven methods have become useful tools to aid in the discovery of new materials for CO\textsubscript{2} capture. For example, surrogate models can be constructed from GCMC-simulated adsorption data to map the structure-property relationship of nanoporous materials, which may then be used to accelerate the HTCS of previously unexplored databases of adsorbents~\cite{simon2015best, bucior2019energy}. Deep generative models can be trained with simulated adsorption data to develop property-orientated generative algorithms to discover new materials on the latent chemical space optimized for gas capture applications by an inverse design approach~\cite{sanchez2018inverse, yao2021inverse, pollice2021data}.
    
    As ML model prediction accuracy and data quality are intrinsically related, there is an apparent requirement to assess the impact of uncertainty from molecular-level simulations on the confidence of machine learning model predictions~\cite{nigam2021assigning}. The efficacy of the surrogate models, for example, depends on the quality of information provided by the material feature vector, and so concerted efforts have been made to develop useful representations of MOFs.~\cite{jablonka2020big} However, little is known about the impact of force field selection on the interpretability of surrogate model predictions. Particularly in the case of MOFs with coordinatively unsaturated metals, different generic force fields and charge assignment schemes can deliver dramatically different results.
    
    Therefore, the importance of material features --- learned either explicitly by the surrogate model as in decision trees, or extracted through feature permutations / SHAP values~\cite{lundberg2018consistent} --- may be subject to considerable discrepancies. Feature importance analyses are useful to guide the design of new functional MOFs, and so it is desirable to understand the differences (if any) that arise from different levels of molecular theory. 

\section*{Code availability}

    
    The Jupyter notebooks providing the \texttt{panel} visualisation of the isotherm curves, enthalpy of adsorption data, and the t-SNE + DBSCAN analysis of the chemical and geometric properties of MOFs are distributed alongside the database in the Zenodo\cite{lopes_oliveira_felipe_2022_7106174} repository.

\bibliography{main}

\begin{thebibliography}{10}
\urlstyle{rm}
\expandafter\ifx\csname url\endcsname\relax
  \def\url#1{\texttt{#1}}\fi
\expandafter\ifx\csname urlprefix\endcsname\relax\def\urlprefix{URL }\fi
\expandafter\ifx\csname doiprefix\endcsname\relax\def\doiprefix{DOI: }\fi
\providecommand{\bibinfo}[2]{#2}
\providecommand{\eprint}[2][]{\url{#2}}

\bibitem{mac2017role}
\bibinfo{author}{Mac~Dowell, N.}, \bibinfo{author}{Fennell, P.~S.},
  \bibinfo{author}{Shah, N.} \& \bibinfo{author}{Maitland, G.~C.}
\newblock \bibinfo{journal}{\bibinfo{title}{The role of co2 capture and
  utilization in mitigating climate change}}.
\newblock {\emph{\JournalTitle{Nature Climate Change}}}
  \textbf{\bibinfo{volume}{7}}, \bibinfo{pages}{243--249}
  (\bibinfo{year}{2017}).

\bibitem{metz2005ipcc}
\bibinfo{author}{Metz, B.}, \bibinfo{author}{Davidson, O.},
  \bibinfo{author}{De~Coninck, H.}, \bibinfo{author}{Loos, M.} \&
  \bibinfo{author}{Meyer, L.}
\newblock \emph{\bibinfo{title}{IPCC special report on carbon dioxide capture
  and storage}} (\bibinfo{publisher}{Cambridge: Cambridge University Press},
  \bibinfo{year}{2005}).

\bibitem{sholl2016seven}
\bibinfo{author}{Sholl, D.~S.} \& \bibinfo{author}{Lively, R.~P.}
\newblock \bibinfo{journal}{\bibinfo{title}{Seven chemical separations to
  change the world}}.
\newblock {\emph{\JournalTitle{Nature}}} \textbf{\bibinfo{volume}{532}},
  \bibinfo{pages}{435--437} (\bibinfo{year}{2016}).

\bibitem{samanta2012post}
\bibinfo{author}{Samanta, A.}, \bibinfo{author}{Zhao, A.},
  \bibinfo{author}{Shimizu, G.~K.}, \bibinfo{author}{Sarkar, P.} \&
  \bibinfo{author}{Gupta, R.}
\newblock \bibinfo{journal}{\bibinfo{title}{Post-combustion co2 capture using
  solid sorbents: a review}}.
\newblock {\emph{\JournalTitle{Industrial \& Engineering Chemistry Research}}}
  \textbf{\bibinfo{volume}{51}}, \bibinfo{pages}{1438--1463}
  (\bibinfo{year}{2012}).

\bibitem{yaghi2003reticular}
\bibinfo{author}{Yaghi, O.~M.} \emph{et~al.}
\newblock \bibinfo{journal}{\bibinfo{title}{Reticular synthesis and the design
  of new materials}}.
\newblock {\emph{\JournalTitle{Nature}}} \textbf{\bibinfo{volume}{423}},
  \bibinfo{pages}{705--714} (\bibinfo{year}{2003}).

\bibitem{furukawa2013chemistry}
\bibinfo{author}{Furukawa, H.}, \bibinfo{author}{Cordova, K.~E.},
  \bibinfo{author}{O’Keeffe, M.} \& \bibinfo{author}{Yaghi, O.~M.}
\newblock \bibinfo{journal}{\bibinfo{title}{The chemistry and applications of
  metal-organic frameworks}}.
\newblock {\emph{\JournalTitle{Science}}} \textbf{\bibinfo{volume}{341}},
  \bibinfo{pages}{1230444} (\bibinfo{year}{2013}).

\bibitem{chen2022reticular}
\bibinfo{author}{Chen, Z.}, \bibinfo{author}{Kirlikovali, K.~O.},
  \bibinfo{author}{Li, P.} \& \bibinfo{author}{Farha, O.~K.}
\newblock \bibinfo{journal}{\bibinfo{title}{Reticular chemistry for highly
  porous metal--organic frameworks: The chemistry and applications}}.
\newblock {\emph{\JournalTitle{Accounts of chemical research}}}
  \textbf{\bibinfo{volume}{55}}, \bibinfo{pages}{579--591}
  (\bibinfo{year}{2022}).

\bibitem{maia2021co}
\bibinfo{author}{Maia, R.~A.}, \bibinfo{author}{Louis, B.},
  \bibinfo{author}{Gao, W.} \& \bibinfo{author}{Wang, Q.}
\newblock \bibinfo{journal}{\bibinfo{title}{Co 2 adsorption mechanisms on mofs:
  a case study of open metal sites, ultra-microporosity and flexible
  framework}}.
\newblock {\emph{\JournalTitle{Reaction Chemistry \& Engineering}}}
  \textbf{\bibinfo{volume}{6}}, \bibinfo{pages}{1118--1133}
  (\bibinfo{year}{2021}).

\bibitem{cote2005porous}
\bibinfo{author}{Cote, A.~P.} \emph{et~al.}
\newblock \bibinfo{journal}{\bibinfo{title}{Porous, crystalline, covalent
  organic frameworks}}.
\newblock {\emph{\JournalTitle{science}}} \textbf{\bibinfo{volume}{310}},
  \bibinfo{pages}{1166--1170} (\bibinfo{year}{2005}).

\bibitem{diercks2017atom}
\bibinfo{author}{Diercks, C.~S.} \& \bibinfo{author}{Yaghi, O.~M.}
\newblock \bibinfo{journal}{\bibinfo{title}{The atom, the molecule, and the
  covalent organic framework}}.
\newblock {\emph{\JournalTitle{Science}}} \textbf{\bibinfo{volume}{355}},
  \bibinfo{pages}{eaal1585} (\bibinfo{year}{2017}).

\bibitem{maia2021co2}
\bibinfo{author}{Maia, R.~A.} \emph{et~al.}
\newblock \bibinfo{journal}{\bibinfo{title}{Co2 capture by hydroxylated
  azine-based covalent organic frameworks}}.
\newblock {\emph{\JournalTitle{Chemistry--A European Journal}}}
  \textbf{\bibinfo{volume}{27}}, \bibinfo{pages}{8048--8055}
  (\bibinfo{year}{2021}).

\bibitem{banerjee2008high}
\bibinfo{author}{Banerjee, R.} \emph{et~al.}
\newblock \bibinfo{journal}{\bibinfo{title}{High-throughput synthesis of
  zeolitic imidazolate frameworks and application to co2 capture}}.
\newblock {\emph{\JournalTitle{Science}}} \textbf{\bibinfo{volume}{319}},
  \bibinfo{pages}{939--943} (\bibinfo{year}{2008}).

\bibitem{yang2017principles}
\bibinfo{author}{Yang, J.} \emph{et~al.}
\newblock \bibinfo{journal}{\bibinfo{title}{Principles of designing extra-large
  pore openings and cages in zeolitic imidazolate frameworks}}.
\newblock {\emph{\JournalTitle{Journal of the American Chemical Society}}}
  \textbf{\bibinfo{volume}{139}}, \bibinfo{pages}{6448--6455}
  (\bibinfo{year}{2017}).

\bibitem{wang2022large}
\bibinfo{author}{Wang, H.}, \bibinfo{author}{Pei, X.},
  \bibinfo{author}{Kalmutzki, M.~J.}, \bibinfo{author}{Yang, J.} \&
  \bibinfo{author}{Yaghi, O.~M.}
\newblock \bibinfo{journal}{\bibinfo{title}{Large cages of zeolitic imidazolate
  frameworks}}.
\newblock {\emph{\JournalTitle{Accounts of Chemical Research}}}
  \textbf{\bibinfo{volume}{55}}, \bibinfo{pages}{707--721}
  (\bibinfo{year}{2022}).

\bibitem{lin2012silico}
\bibinfo{author}{Lin, L.-C.} \emph{et~al.}
\newblock \bibinfo{journal}{\bibinfo{title}{In silico screening of
  carbon-capture materials}}.
\newblock {\emph{\JournalTitle{Nature materials}}}
  \textbf{\bibinfo{volume}{11}}, \bibinfo{pages}{633--641}
  (\bibinfo{year}{2012}).

\bibitem{chung2014computation}
\bibinfo{author}{Chung, Y.~G.} \emph{et~al.}
\newblock \bibinfo{journal}{\bibinfo{title}{Computation-ready, experimental
  metal--organic frameworks: A tool to enable high-throughput screening of
  nanoporous crystals}}.
\newblock {\emph{\JournalTitle{Chemistry of Materials}}}
  \textbf{\bibinfo{volume}{26}}, \bibinfo{pages}{6185--6192}
  (\bibinfo{year}{2014}).

\bibitem{tong2017exploring}
\bibinfo{author}{Tong, M.}, \bibinfo{author}{Lan, Y.}, \bibinfo{author}{Yang,
  Q.} \& \bibinfo{author}{Zhong, C.}
\newblock \bibinfo{journal}{\bibinfo{title}{Exploring the structure-property
  relationships of covalent organic frameworks for noble gas separations}}.
\newblock {\emph{\JournalTitle{Chemical Engineering Science}}}
  \textbf{\bibinfo{volume}{168}}, \bibinfo{pages}{456--464}
  (\bibinfo{year}{2017}).

\bibitem{moghadam2017development}
\bibinfo{author}{Moghadam, P.~Z.} \emph{et~al.}
\newblock \bibinfo{journal}{\bibinfo{title}{Development of a cambridge
  structural database subset: a collection of metal--organic frameworks for
  past, present, and future}}.
\newblock {\emph{\JournalTitle{Chemistry of Materials}}}
  \textbf{\bibinfo{volume}{29}}, \bibinfo{pages}{2618--2625}
  (\bibinfo{year}{2017}).

\bibitem{chung2019advances}
\bibinfo{author}{Chung, Y.~G.} \emph{et~al.}
\newblock \bibinfo{journal}{\bibinfo{title}{Advances, updates, and analytics
  for the computation-ready, experimental metal--organic framework database:
  Core mof 2019}}.
\newblock {\emph{\JournalTitle{Journal of Chemical \& Engineering Data}}}
  \textbf{\bibinfo{volume}{64}}, \bibinfo{pages}{5985--5998}
  (\bibinfo{year}{2019}).

\bibitem{ongari2019building}
\bibinfo{author}{Ongari, D.}, \bibinfo{author}{Yakutovich, A.~V.},
  \bibinfo{author}{Talirz, L.} \& \bibinfo{author}{Smit, B.}
\newblock \bibinfo{journal}{\bibinfo{title}{Building a consistent and
  reproducible database for adsorption evaluation in covalent--organic
  frameworks}}.
\newblock {\emph{\JournalTitle{ACS central science}}}
  \textbf{\bibinfo{volume}{5}}, \bibinfo{pages}{1663--1675}
  (\bibinfo{year}{2019}).

\bibitem{colon2017topologically}
\bibinfo{author}{Col{\'o}n, Y.~J.}, \bibinfo{author}{Gomez-Gualdron, D.~A.} \&
  \bibinfo{author}{Snurr, R.~Q.}
\newblock \bibinfo{journal}{\bibinfo{title}{Topologically guided, automated
  construction of metal--organic frameworks and their evaluation for
  energy-related applications}}.
\newblock {\emph{\JournalTitle{Crystal Growth \& Design}}}
  \textbf{\bibinfo{volume}{17}}, \bibinfo{pages}{5801--5810}
  (\bibinfo{year}{2017}).

\bibitem{rosen2021machine}
\bibinfo{author}{Rosen, A.~S.} \emph{et~al.}
\newblock \bibinfo{journal}{\bibinfo{title}{Machine learning the
  quantum-chemical properties of metal--organic frameworks for accelerated
  materials discovery}}.
\newblock {\emph{\JournalTitle{Matter}}} \textbf{\bibinfo{volume}{4}},
  \bibinfo{pages}{1578--1597} (\bibinfo{year}{2021}).

\bibitem{lyu2020digital}
\bibinfo{author}{Lyu, H.}, \bibinfo{author}{Ji, Z.}, \bibinfo{author}{Wuttke,
  S.} \& \bibinfo{author}{Yaghi, O.~M.}
\newblock \bibinfo{journal}{\bibinfo{title}{Digital reticular chemistry}}.
\newblock {\emph{\JournalTitle{Chem}}} \textbf{\bibinfo{volume}{6}},
  \bibinfo{pages}{2219--2241} (\bibinfo{year}{2020}).

\bibitem{ji2021molecules}
\bibinfo{author}{Ji, Z.} \emph{et~al.}
\newblock \bibinfo{journal}{\bibinfo{title}{From molecules to frameworks to
  superframework crystals}}.
\newblock {\emph{\JournalTitle{Advanced Materials}}}
  \textbf{\bibinfo{volume}{33}}, \bibinfo{pages}{2103808}
  (\bibinfo{year}{2021}).

\bibitem{colon2014high}
\bibinfo{author}{Col{\'o}n, Y.~J.} \& \bibinfo{author}{Snurr, R.~Q.}
\newblock \bibinfo{journal}{\bibinfo{title}{High-throughput computational
  screening of metal--organic frameworks}}.
\newblock {\emph{\JournalTitle{Chemical Society Reviews}}}
  \textbf{\bibinfo{volume}{43}}, \bibinfo{pages}{5735--5749}
  (\bibinfo{year}{2014}).

\bibitem{majumdar2021diversifying}
\bibinfo{author}{Majumdar, S.}, \bibinfo{author}{Moosavi, S.~M.},
  \bibinfo{author}{Jablonka, K.~M.}, \bibinfo{author}{Ongari, D.} \&
  \bibinfo{author}{Smit, B.}
\newblock \bibinfo{journal}{\bibinfo{title}{Diversifying databases of metal
  organic frameworks for high-throughput computational screening}}.
\newblock {\emph{\JournalTitle{ACS applied materials \& interfaces}}}
  \textbf{\bibinfo{volume}{13}}, \bibinfo{pages}{61004--61014}
  (\bibinfo{year}{2021}).

\bibitem{dubbeldam2019design}
\bibinfo{author}{Dubbeldam, D.}, \bibinfo{author}{Walton, K.~S.},
  \bibinfo{author}{Vlugt, T.~J.} \& \bibinfo{author}{Calero, S.}
\newblock \bibinfo{journal}{\bibinfo{title}{Design, parameterization, and
  implementation of atomic force fields for adsorption in nanoporous
  materials}}.
\newblock {\emph{\JournalTitle{Advanced Theory and Simulations}}}
  \textbf{\bibinfo{volume}{2}}, \bibinfo{pages}{1900135}
  (\bibinfo{year}{2019}).

\bibitem{rappe1992uff}
\bibinfo{author}{Rapp{\'e}, A.~K.}, \bibinfo{author}{Casewit, C.~J.},
  \bibinfo{author}{Colwell, K.}, \bibinfo{author}{Goddard~III, W.~A.} \&
  \bibinfo{author}{Skiff, W.~M.}
\newblock \bibinfo{journal}{\bibinfo{title}{Uff, a full periodic table force
  field for molecular mechanics and molecular dynamics simulations}}.
\newblock {\emph{\JournalTitle{Journal of the American chemical society}}}
  \textbf{\bibinfo{volume}{114}}, \bibinfo{pages}{10024--10035}
  (\bibinfo{year}{1992}).

\bibitem{mayo1990dreiding}
\bibinfo{author}{Mayo, S.~L.}, \bibinfo{author}{Olafson, B.~D.} \&
  \bibinfo{author}{Goddard, W.~A.}
\newblock \bibinfo{journal}{\bibinfo{title}{Dreiding: a generic force field for
  molecular simulations}}.
\newblock {\emph{\JournalTitle{Journal of Physical chemistry}}}
  \textbf{\bibinfo{volume}{94}}, \bibinfo{pages}{8897--8909}
  (\bibinfo{year}{1990}).

\bibitem{potoff2001vapor}
\bibinfo{author}{Potoff, J.~J.} \& \bibinfo{author}{Siepmann, J.~I.}
\newblock \bibinfo{journal}{\bibinfo{title}{Vapor--liquid equilibria of
  mixtures containing alkanes, carbon dioxide, and nitrogen}}.
\newblock {\emph{\JournalTitle{AIChE journal}}} \textbf{\bibinfo{volume}{47}},
  \bibinfo{pages}{1676--1682} (\bibinfo{year}{2001}).

\bibitem{allen2017computer}
\bibinfo{author}{Allen, M.~P.} \& \bibinfo{author}{Tildesley, D.~J.}
\newblock \emph{\bibinfo{title}{Computer simulation of liquids}}
  (\bibinfo{publisher}{Oxford university press}, \bibinfo{year}{2017}).

\bibitem{jorgensen1986optimized}
\bibinfo{author}{Jorgensen, W.~L.}
\newblock \bibinfo{journal}{\bibinfo{title}{Optimized intermolecular potential
  functions for liquid alcohols}}.
\newblock {\emph{\JournalTitle{The Journal of Physical Chemistry}}}
  \textbf{\bibinfo{volume}{90}}, \bibinfo{pages}{1276--1284}
  (\bibinfo{year}{1986}).

\bibitem{bayly1993well}
\bibinfo{author}{Bayly, C.~I.}, \bibinfo{author}{Cieplak, P.},
  \bibinfo{author}{Cornell, W.} \& \bibinfo{author}{Kollman, P.~A.}
\newblock \bibinfo{journal}{\bibinfo{title}{A well-behaved electrostatic
  potential based method using charge restraints for deriving atomic charges:
  the resp model}}.
\newblock {\emph{\JournalTitle{The Journal of Physical Chemistry}}}
  \textbf{\bibinfo{volume}{97}}, \bibinfo{pages}{10269--10280}
  (\bibinfo{year}{1993}).

\bibitem{breneman1990determining}
\bibinfo{author}{Breneman, C.~M.} \& \bibinfo{author}{Wiberg, K.~B.}
\newblock \bibinfo{journal}{\bibinfo{title}{Determining atom-centered monopoles
  from molecular electrostatic potentials. the need for high sampling density
  in formamide conformational analysis}}.
\newblock {\emph{\JournalTitle{Journal of Computational Chemistry}}}
  \textbf{\bibinfo{volume}{11}}, \bibinfo{pages}{361--373}
  (\bibinfo{year}{1990}).

\bibitem{campana2009electrostatic}
\bibinfo{author}{Campa{\~n}{\'a}, C.}, \bibinfo{author}{Mussard, B.} \&
  \bibinfo{author}{Woo, T.~K.}
\newblock \bibinfo{journal}{\bibinfo{title}{Electrostatic potential derived
  atomic charges for periodic systems using a modified error functional}}.
\newblock {\emph{\JournalTitle{Journal of Chemical Theory and Computation}}}
  \textbf{\bibinfo{volume}{5}}, \bibinfo{pages}{2866--2878}
  (\bibinfo{year}{2009}).

\bibitem{manz2016introducing}
\bibinfo{author}{Manz, T.~A.} \& \bibinfo{author}{Limas, N.~G.}
\newblock \bibinfo{journal}{\bibinfo{title}{Introducing ddec6 atomic population
  analysis: part 1. charge partitioning theory and methodology}}.
\newblock {\emph{\JournalTitle{RSC advances}}} \textbf{\bibinfo{volume}{6}},
  \bibinfo{pages}{47771--47801} (\bibinfo{year}{2016}).

\bibitem{limas2016introducing}
\bibinfo{author}{Limas, N.~G.} \& \bibinfo{author}{Manz, T.~A.}
\newblock \bibinfo{journal}{\bibinfo{title}{Introducing ddec6 atomic population
  analysis: part 2. computed results for a wide range of periodic and
  nonperiodic materials}}.
\newblock {\emph{\JournalTitle{RSC advances}}} \textbf{\bibinfo{volume}{6}},
  \bibinfo{pages}{45727--45747} (\bibinfo{year}{2016}).

\bibitem{manz2017introducing}
\bibinfo{author}{Manz, T.~A.}
\newblock \bibinfo{journal}{\bibinfo{title}{Introducing ddec6 atomic population
  analysis: part 3. comprehensive method to compute bond orders}}.
\newblock {\emph{\JournalTitle{RSC advances}}} \textbf{\bibinfo{volume}{7}},
  \bibinfo{pages}{45552--45581} (\bibinfo{year}{2017}).

\bibitem{rappe1991charge}
\bibinfo{author}{Rappe, A.~K.} \& \bibinfo{author}{Goddard~III, W.~A.}
\newblock \bibinfo{journal}{\bibinfo{title}{Charge equilibration for molecular
  dynamics simulations}}.
\newblock {\emph{\JournalTitle{The Journal of Physical Chemistry}}}
  \textbf{\bibinfo{volume}{95}}, \bibinfo{pages}{3358--3363}
  (\bibinfo{year}{1991}).

\bibitem{ramachandran1996toward}
\bibinfo{author}{Ramachandran, S.}, \bibinfo{author}{Lenz, T.},
  \bibinfo{author}{Skiff, W.} \& \bibinfo{author}{Rapp{\'e}, A.}
\newblock \bibinfo{journal}{\bibinfo{title}{Toward an understanding of zeolite
  y as a cracking catalyst with the use of periodic charge equilibration}}.
\newblock {\emph{\JournalTitle{The Journal of Physical Chemistry}}}
  \textbf{\bibinfo{volume}{100}}, \bibinfo{pages}{5898--5907}
  (\bibinfo{year}{1996}).

\bibitem{wilmer2012extended}
\bibinfo{author}{Wilmer, C.~E.}, \bibinfo{author}{Kim, K.~C.} \&
  \bibinfo{author}{Snurr, R.~Q.}
\newblock \bibinfo{journal}{\bibinfo{title}{An extended charge equilibration
  method}}.
\newblock {\emph{\JournalTitle{The journal of physical chemistry letters}}}
  \textbf{\bibinfo{volume}{3}}, \bibinfo{pages}{2506--2511}
  (\bibinfo{year}{2012}).

\bibitem{wells2015charge}
\bibinfo{author}{Wells, B.~A.}, \bibinfo{author}{De~Bruin-Dickason, C.} \&
  \bibinfo{author}{Chaffee, A.~L.}
\newblock \bibinfo{journal}{\bibinfo{title}{Charge equilibration based on
  atomic ionization in metal--organic frameworks}}.
\newblock {\emph{\JournalTitle{The Journal of Physical Chemistry C}}}
  \textbf{\bibinfo{volume}{119}}, \bibinfo{pages}{456--466}
  (\bibinfo{year}{2015}).

\bibitem{altintas2020role}
\bibinfo{author}{Altintas, C.} \& \bibinfo{author}{Keskin, S.}
\newblock \bibinfo{journal}{\bibinfo{title}{Role of partial charge assignment
  methods in high-throughput screening of mof adsorbents and membranes for co
  2/ch 4 separation}}.
\newblock {\emph{\JournalTitle{Molecular Systems Design \& Engineering}}}
  \textbf{\bibinfo{volume}{5}}, \bibinfo{pages}{532--543}
  (\bibinfo{year}{2020}).

\bibitem{liu2022benchmarking}
\bibinfo{author}{Liu, S.} \& \bibinfo{author}{Luan, B.}
\newblock \bibinfo{journal}{\bibinfo{title}{Benchmarking various types of
  partial atomic charges for classical all-atom simulations of metal-organic
  frameworks}}.
\newblock {\emph{\JournalTitle{Nanoscale}}}  (\bibinfo{year}{2022}).

\bibitem{zheng2009computational}
\bibinfo{author}{Zheng, C.}, \bibinfo{author}{Liu, D.}, \bibinfo{author}{Yang,
  Q.}, \bibinfo{author}{Zhong, C.} \& \bibinfo{author}{Mi, J.}
\newblock \bibinfo{journal}{\bibinfo{title}{Computational study on the
  influences of framework charges on co2 uptake in metal- organic frameworks}}.
\newblock {\emph{\JournalTitle{Industrial \& engineering chemistry research}}}
  \textbf{\bibinfo{volume}{48}}, \bibinfo{pages}{10479--10484}
  (\bibinfo{year}{2009}).

\bibitem{hamad2015atomic}
\bibinfo{author}{Hamad, S.}, \bibinfo{author}{Balestra, S.~R.},
  \bibinfo{author}{Bueno-Perez, R.}, \bibinfo{author}{Calero, S.} \&
  \bibinfo{author}{Ruiz-Salvador, A.~R.}
\newblock \bibinfo{journal}{\bibinfo{title}{Atomic charges for modeling
  metal--organic frameworks: Why and how}}.
\newblock {\emph{\JournalTitle{Journal of Solid State Chemistry}}}
  \textbf{\bibinfo{volume}{223}}, \bibinfo{pages}{144--151}
  (\bibinfo{year}{2015}).

\bibitem{ongari2018evaluating}
\bibinfo{author}{Ongari, D.} \emph{et~al.}
\newblock \bibinfo{journal}{\bibinfo{title}{Evaluating charge equilibration
  methods to generate electrostatic fields in nanoporous materials}}.
\newblock {\emph{\JournalTitle{Journal of chemical theory and computation}}}
  \textbf{\bibinfo{volume}{15}}, \bibinfo{pages}{382--401}
  (\bibinfo{year}{2018}).

\bibitem{sladekova2020effect}
\bibinfo{author}{Sladekova, K.} \emph{et~al.}
\newblock \bibinfo{journal}{\bibinfo{title}{The effect of atomic point charges
  on adsorption isotherms of co2 and water in metal organic frameworks}}.
\newblock {\emph{\JournalTitle{Adsorption}}} \textbf{\bibinfo{volume}{26}},
  \bibinfo{pages}{663--685} (\bibinfo{year}{2020}).

\bibitem{gasteiger1980iterative}
\bibinfo{author}{Gasteiger, J.} \& \bibinfo{author}{Marsili, M.}
\newblock \bibinfo{journal}{\bibinfo{title}{Iterative partial equalization of
  orbital electronegativity—a rapid access to atomic charges}}.
\newblock {\emph{\JournalTitle{Tetrahedron}}} \textbf{\bibinfo{volume}{36}},
  \bibinfo{pages}{3219--3228} (\bibinfo{year}{1980}).

\bibitem{eqeq}
\bibinfo{author}{Ongari, D.}
\newblock \bibinfo{title}{{Charge equilibration method for crystal structures
  Software}}.
\newblock \bibinfo{howpublished}{\url{https://github.com/danieleongari/EQeq}}
  (\bibinfo{year}{2020}).
\newblock \bibinfo{note}{[Online; accessed 11-May-2022]}.

\bibitem{dubbeldam2016raspa}
\bibinfo{author}{Dubbeldam, D.}, \bibinfo{author}{Calero, S.},
  \bibinfo{author}{Ellis, D.~E.} \& \bibinfo{author}{Snurr, R.~Q.}
\newblock \bibinfo{journal}{\bibinfo{title}{Raspa: molecular simulation
  software for adsorption and diffusion in flexible nanoporous materials}}.
\newblock {\emph{\JournalTitle{Molecular Simulation}}}
  \textbf{\bibinfo{volume}{42}}, \bibinfo{pages}{81--101}
  (\bibinfo{year}{2016}).

\bibitem{dubbeldam2013inner}
\bibinfo{author}{Dubbeldam, D.}, \bibinfo{author}{Torres-Knoop, A.} \&
  \bibinfo{author}{Walton, K.~S.}
\newblock \bibinfo{journal}{\bibinfo{title}{On the inner workings of monte
  carlo codes}}.
\newblock {\emph{\JournalTitle{Molecular Simulation}}}
  \textbf{\bibinfo{volume}{39}}, \bibinfo{pages}{1253--1292}
  (\bibinfo{year}{2013}).

\bibitem{peng1976industrial}
\bibinfo{author}{Peng, D.} \& \bibinfo{author}{Robinson, D.}
\newblock \bibinfo{journal}{\bibinfo{title}{Industrial engineering chemistry
  fundamentals}}.
\newblock {\emph{\JournalTitle{A New Two-Constant Equation of State}}}
  \textbf{\bibinfo{volume}{15}}, \bibinfo{pages}{59--64}
  (\bibinfo{year}{1976}).

\bibitem{vlugt2008computing}
\bibinfo{author}{Vlugt, T.}, \bibinfo{author}{Garc{\'\i}a-P{\'e}rez, E.},
  \bibinfo{author}{Dubbeldam, D.}, \bibinfo{author}{Ban, S.} \&
  \bibinfo{author}{Calero, S.}
\newblock \bibinfo{journal}{\bibinfo{title}{Computing the heat of adsorption
  using molecular simulations: the effect of strong coulombic interactions}}.
\newblock {\emph{\JournalTitle{Journal of chemical theory and computation}}}
  \textbf{\bibinfo{volume}{4}}, \bibinfo{pages}{1107--1118}
  (\bibinfo{year}{2008}).

\bibitem{white1997effective}
\bibinfo{author}{White~Jr, K.~P.}
\newblock \bibinfo{journal}{\bibinfo{title}{An effective truncation heuristic
  for bias reduction in simulation output}}.
\newblock {\emph{\JournalTitle{Simulation}}} \textbf{\bibinfo{volume}{69}},
  \bibinfo{pages}{323--334} (\bibinfo{year}{1997}).

\bibitem{pymser}
\bibinfo{author}{Oliveira, F.~L.} \& \bibinfo{author}{Ferreira, R. N.~B.}
\newblock \bibinfo{title}{pymser: A python library to apply the marginal
  standard error rule (mser) for transient regime detection and truncation on
  grand canonical monte carlo adsorption simulations.}
\newblock \bibinfo{howpublished}{\url{https://github.com/IBM/pymser}}
  (\bibinfo{year}{2022}).
\newblock \bibinfo{note}{[Online; accessed 10-Aug-2022]}.

\bibitem{harris2020array}
\bibinfo{author}{Harris, C.~R.} \emph{et~al.}
\newblock \bibinfo{journal}{\bibinfo{title}{Array programming with numpy}}.
\newblock {\emph{\JournalTitle{Nature}}} \textbf{\bibinfo{volume}{585}},
  \bibinfo{pages}{357--362} (\bibinfo{year}{2020}).

\bibitem{ioannidis2016molsimplify}
\bibinfo{author}{Ioannidis, E.~I.}, \bibinfo{author}{Gani, T.~Z.} \&
  \bibinfo{author}{Kulik, H.~J.}
\newblock \bibinfo{title}{molsimplify: A toolkit for automating discovery in
  inorganic chemistry} (\bibinfo{year}{2016}).

\bibitem{van2008visualizing}
\bibinfo{author}{Van~der Maaten, L.} \& \bibinfo{author}{Hinton, G.}
\newblock \bibinfo{journal}{\bibinfo{title}{Visualizing data using t-sne.}}
\newblock {\emph{\JournalTitle{Journal of machine learning research}}}
  \textbf{\bibinfo{volume}{9}} (\bibinfo{year}{2008}).

\bibitem{willems2012algorithms}
\bibinfo{author}{Willems, T.~F.}, \bibinfo{author}{Rycroft, C.~H.},
  \bibinfo{author}{Kazi, M.}, \bibinfo{author}{Meza, J.~C.} \&
  \bibinfo{author}{Haranczyk, M.}
\newblock \bibinfo{journal}{\bibinfo{title}{Algorithms and tools for
  high-throughput geometry-based analysis of crystalline porous materials}}.
\newblock {\emph{\JournalTitle{Microporous and Mesoporous Materials}}}
  \textbf{\bibinfo{volume}{149}}, \bibinfo{pages}{134--141}
  (\bibinfo{year}{2012}).

\bibitem{bi2012dbscan}
\bibinfo{author}{Bi, F.}, \bibinfo{author}{Wang, W.} \& \bibinfo{author}{Chen,
  L.}
\newblock \bibinfo{journal}{\bibinfo{title}{Dbscan: density-based spatial
  clustering of applications with noise}}.
\newblock {\emph{\JournalTitle{J. Nanjing Univ}}}
  \textbf{\bibinfo{volume}{48}}, \bibinfo{pages}{491--498}
  (\bibinfo{year}{2012}).

\bibitem{janet2017resolving}
\bibinfo{author}{Janet, J.~P.} \& \bibinfo{author}{Kulik, H.~J.}
\newblock \bibinfo{journal}{\bibinfo{title}{Resolving transition metal chemical
  space: Feature selection for machine learning and structure--property
  relationships}}.
\newblock {\emph{\JournalTitle{The Journal of Physical Chemistry A}}}
  \textbf{\bibinfo{volume}{121}}, \bibinfo{pages}{8939--8954}
  (\bibinfo{year}{2017}).

\bibitem{nandy2018strategies}
\bibinfo{author}{Nandy, A.}, \bibinfo{author}{Duan, C.},
  \bibinfo{author}{Janet, J.~P.}, \bibinfo{author}{Gugler, S.} \&
  \bibinfo{author}{Kulik, H.~J.}
\newblock \bibinfo{journal}{\bibinfo{title}{Strategies and software for machine
  learning accelerated discovery in transition metal chemistry}}.
\newblock {\emph{\JournalTitle{Industrial \& Engineering Chemistry Research}}}
  \textbf{\bibinfo{volume}{57}}, \bibinfo{pages}{13973--13986}
  (\bibinfo{year}{2018}).

\bibitem{moosavi2020understanding}
\bibinfo{author}{Moosavi, S.~M.} \emph{et~al.}
\newblock \bibinfo{journal}{\bibinfo{title}{Understanding the diversity of the
  metal-organic framework ecosystem}}.
\newblock {\emph{\JournalTitle{Nature communications}}}
  \textbf{\bibinfo{volume}{11}}, \bibinfo{pages}{1--10} (\bibinfo{year}{2020}).

\bibitem{park2019well}
\bibinfo{author}{Park, J.} \emph{et~al.}
\newblock \bibinfo{journal}{\bibinfo{title}{How well do approximate models of
  adsorption-based co2 capture processes predict results of detailed process
  models?}}
\newblock {\emph{\JournalTitle{Industrial \& Engineering Chemistry Research}}}
  \textbf{\bibinfo{volume}{59}}, \bibinfo{pages}{7097--7108}
  (\bibinfo{year}{2019}).

\bibitem{lopes_oliveira_felipe_2022_7106174}
\bibinfo{author}{Oliveira, F.~L.} \emph{et~al.}
\newblock \bibinfo{title}{{CRAFTED - An exploratory database of simulated
  adsorption isotherms of metal-organic frameworks}},
  \url{10.5281/zenodo.7106174} (\bibinfo{year}{2022}).

\bibitem{simon2015best}
\bibinfo{author}{Simon, C.~M.}, \bibinfo{author}{Mercado, R.},
  \bibinfo{author}{Schnell, S.~K.}, \bibinfo{author}{Smit, B.} \&
  \bibinfo{author}{Haranczyk, M.}
\newblock \bibinfo{journal}{\bibinfo{title}{What are the best materials to
  separate a xenon/krypton mixture?}}
\newblock {\emph{\JournalTitle{Chemistry of Materials}}}
  \textbf{\bibinfo{volume}{27}}, \bibinfo{pages}{4459--4475}
  (\bibinfo{year}{2015}).

\bibitem{bucior2019energy}
\bibinfo{author}{Bucior, B.~J.} \emph{et~al.}
\newblock \bibinfo{journal}{\bibinfo{title}{Energy-based descriptors to rapidly
  predict hydrogen storage in metal--organic frameworks}}.
\newblock {\emph{\JournalTitle{Molecular Systems Design \& Engineering}}}
  \textbf{\bibinfo{volume}{4}}, \bibinfo{pages}{162--174}
  (\bibinfo{year}{2019}).

\bibitem{sanchez2018inverse}
\bibinfo{author}{Sanchez-Lengeling, B.} \& \bibinfo{author}{Aspuru-Guzik, A.}
\newblock \bibinfo{journal}{\bibinfo{title}{Inverse molecular design using
  machine learning: Generative models for matter engineering}}.
\newblock {\emph{\JournalTitle{Science}}} \textbf{\bibinfo{volume}{361}},
  \bibinfo{pages}{360--365} (\bibinfo{year}{2018}).

\bibitem{yao2021inverse}
\bibinfo{author}{Yao, Z.} \emph{et~al.}
\newblock \bibinfo{journal}{\bibinfo{title}{Inverse design of nanoporous
  crystalline reticular materials with deep generative models}}.
\newblock {\emph{\JournalTitle{Nature Machine Intelligence}}}
  \textbf{\bibinfo{volume}{3}}, \bibinfo{pages}{76--86} (\bibinfo{year}{2021}).

\bibitem{pollice2021data}
\bibinfo{author}{Pollice, R.} \emph{et~al.}
\newblock \bibinfo{journal}{\bibinfo{title}{Data-driven strategies for
  accelerated materials design}}.
\newblock {\emph{\JournalTitle{Accounts of Chemical Research}}}
  \textbf{\bibinfo{volume}{54}}, \bibinfo{pages}{849--860}
  (\bibinfo{year}{2021}).

\bibitem{nigam2021assigning}
\bibinfo{author}{Nigam, A.} \emph{et~al.}
\newblock \bibinfo{journal}{\bibinfo{title}{Assigning confidence to molecular
  property prediction}}.
\newblock {\emph{\JournalTitle{Expert opinion on drug discovery}}}
  \textbf{\bibinfo{volume}{16}}, \bibinfo{pages}{1009--1023}
  (\bibinfo{year}{2021}).

\bibitem{jablonka2020big}
\bibinfo{author}{Jablonka, K.~M.}, \bibinfo{author}{Ongari, D.},
  \bibinfo{author}{Moosavi, S.~M.} \& \bibinfo{author}{Smit, B.}
\newblock \bibinfo{journal}{\bibinfo{title}{Big-data science in porous
  materials: materials genomics and machine learning}}.
\newblock {\emph{\JournalTitle{Chemical reviews}}}
  \textbf{\bibinfo{volume}{120}}, \bibinfo{pages}{8066--8129}
  (\bibinfo{year}{2020}).

\bibitem{lundberg2018consistent}
\bibinfo{author}{Lundberg, S.~M.}, \bibinfo{author}{Erion, G.~G.} \&
  \bibinfo{author}{Lee, S.-I.}
\newblock \bibinfo{journal}{\bibinfo{title}{Consistent individualized feature
  attribution for tree ensembles}}.
\newblock {\emph{\JournalTitle{arXiv preprint arXiv:1802.03888}}}
  (\bibinfo{year}{2018}).

\end{thebibliography}


\section*{Acknowledgements}

The authors would like to acknowledge Flor Siperstein and Joseph Manning (University of Manchester) for fruitful discussions that helped shape this work.

\section*{Author contributions statement}

F.L.O. developed the GCMC simulation workflow, analyzed the data, compiled the database and wrote the manuscript. C.C. wrote the manuscript and proposed the generation of the database. R.N.B.F. developed the GCMC simulation workflow and wrote the manuscript. B.L. developed the GCMC simulation workflow. A.H.F. proposed the generation of the database. L.S. and M.S. wrote the manuscript. All authors reviewed the manuscript. 

\section*{Competing interests}

The authors declare no competing interests.

\section*{Figures \& Tables}

\begin{table}[ht]
    \centering
    \begin{tabular}{|l|ll|ll|}
    \hline
    \multicolumn{1}{|c|}{\multirow{2}{*}{Atom type}} &
      \multicolumn{2}{|c|}{UFF} &
      \multicolumn{2}{|c|}{DREIDING} \\ \cmidrule(l){2-5} 
    \multicolumn{1}{|c|}{} &   \multicolumn{1}{c|}{$\sigma$ (\AA{})} &  \multicolumn{1}{c|}{$\varepsilon$ (K)} &
      \multicolumn{1}{c|}{$\sigma$ (\AA{})} &  \multicolumn{1}{c|}{$\varepsilon$ (K)} \\ \hline
            H  & 2.571 & 22.14  & 2.84642 & 7.64936   \\
            B  & 3.638 & 90.51  & 3.58141 & 47.80852  \\
            C  & 3.431 & 52.8   & 3.47299 & 47.85884  \\
            N  & 3.261 & 34.7   & 3.26256 & 38.95136  \\
            O  & 3.118 & 30.2   & 3.03315 & 48.16079  \\
            F  & 2.997 & 25.14  & 3.0932  & 36.48545  \\
            Cl & 3.516 & 114.15 & 3.51932 & 142.57003 \\
            Br & 3.732 & 126.22 & 3.51905 & 186.20159 \\
            Al & 4.008 & 253.94 & 3.91105 & 156.00674 \\
            Si & 3.826 & 202.15 & 3.80414 & 156.00674 \\
            P  & 3.695 & 153.37 & 3.69723 & 161.03921 \\
            S  & 3.595 & 137.78 & 3.59032 & 173.11715 \\
            Ga & 3.905 & 208.69 & 3.91105 & 201.29901 \\
            Ge & 3.813 & 190.58 & 3.80414 & 201.29901 \\
            As & 3.769 & 155.38 & 3.69723 & 206.33149 \\
            Se & 3.746 & 146.33 & 3.59032 & 216.39644 \\
            In & 3.976 & 301.21 & 4.08923 & 276.78614 \\
            Sn & 3.913 & 285.12 & 3.98232 & 276.78614 \\
            Sb & 3.938 & 225.78 & 3.87541 & 276.78614 \\
            Te & 3.982 & 200.14 & 3.7685  & 286.85109 \\
            Na & 2.658 & 15.09  & 2.80099 & 251.62377 \\
            Ca & 3.028 & 119.68 & 3.0932  & 25.16238  \\
            Fe & 2.594 & 6.54   & 4.04468 & 27.67861  \\
            Zn & 2.462 & 62.38  & 4.04468 & 27.67861  \\
            Ti & 2.829 & 8.55   & 4.04468 & 27.67861  \\
            Tc & 2.671 & 24.14  & 4.04468 & 27.67861  \\
            Ru & 2.64  & 28.16  & 4.04468 & 27.67861  \\ \hline
    \end{tabular}
    \caption{\label{tab:forcefields} Lennard-Jones parameters for UFF and DREIDING force fields.}
\end{table}

\begin{table}[ht]
\centering
    \begin{tabular}{|c|c|c|c|}
    \hline
    Atom type & $\sigma$ (\AA{}) & $\varepsilon$ (K) & Charge \\ \hline
    C\_co2     & 2.80       & 27.0          & 0.700  \\
    O\_co2     & 3.05       & 79.0          & -0.350 \\
    N\_n2      & 3.31       & 36.0          & -0.482 \\
    N\_com     & -          & -             & 0.964  \\ \hline
    \end{tabular}
    \caption{\label{tab:trappe} Lennard-Jones parameters for TraPPE force field.}
\end{table}

\begin{table}[ht]
    \centering
    \begin{tabular}{|c|c|c|c|}
    \hline
        Gas & Critical temperature (K) & Critical Pressure (Pa) & Acentric factor \\ \hline
        CO\textsubscript{2} & 304.1282                 & 7377300.0              & 0.22394         \\
        N\textsubscript{2}  & 126.192                  & 3395800.0              & 0.0372          \\ \hline
    \end{tabular}
     \caption{\label{tab:critical} Critical parameters for CO\textsubscript{2} and N\textsubscript{2}.}
\end{table}

\end{document}